\documentclass[aps,prb,reprint,showpacs,floatfix,citeautoscript,longbibliography,superscriptaddress]{revtex4-1}
\usepackage{graphicx}
\usepackage{bm,amsmath,amssymb,mathrsfs,dcolumn}
\usepackage[colorlinks=true, linkcolor=blue, citecolor=blue, urlcolor=blue, linktoc=page, bookmarks=false]{hyperref} 
\usepackage[usenames]{color}
\hypersetup{pdfborder=1 1 1,colorlinks=true,citecolor=blue,linkcolor=blue}
\usepackage{epstopdf}
\usepackage{subfigure}
\usepackage{units}
\usepackage{breakurl}
\usepackage{multirow}
\usepackage{graphicx}
\usepackage{booktabs}

\begin{document}

\title{Brownian Motion and Entropic Torque Driven Motion of Domain-Wall in Antiferromagnets}
\author{Zhengren Yan}
\affiliation{Institute for Advanced Materials, South China Academy of Advanced Optoelectronics and Guangdong Provincial Key Laboratory of Quantum Engineering and Quantum Materials, South China Normal University, Guangzhou 510006, China}

\author{Zhiyuan Chen}
\affiliation{Institute for Advanced Materials, South China Academy of Advanced Optoelectronics and Guangdong Provincial Key Laboratory of Quantum Engineering and Quantum Materials, South China Normal University, Guangzhou 510006, China}

\author{Minghui Qin}
\altaffiliation{qinmh@scnu.edu.cn}
\affiliation{Institute for Advanced Materials, South China Academy of Advanced Optoelectronics and Guangdong Provincial Key Laboratory of Quantum Engineering and Quantum Materials, South China Normal University, Guangzhou 510006, China}

\author{Xubing Lu}
\affiliation{Institute for Advanced Materials, South China Academy of Advanced Optoelectronics and Guangdong Provincial Key Laboratory of Quantum Engineering and Quantum Materials, South China Normal University, Guangzhou 510006, China}

\author{Xingsen Gao}
\affiliation{Institute for Advanced Materials, South China Academy of Advanced Optoelectronics and Guangdong Provincial Key Laboratory of Quantum Engineering and Quantum Materials, South China Normal University, Guangzhou 510006, China}

\author{Junming Liu}
\altaffiliation{liujm@nju.edu.cn}
\affiliation{Laboratory of Solid State Microstructures, Nanjing University, Nanjing 210093, China}

\date{\today }

\begin{abstract}
We study the spin dynamics in antiferromagnetic nanowire under an applied temperature gradient using micromagnetic simulations on a classical spin model with an uniaxial anisotropy. The entropic torque driven domain-wall motion and the Brownian motion are discussed in detail, and their competition determines the antiferromagnetic wall motion towards the hotter or colder region. Furthermore, the spin dynamics in an antiferromagnet can be well tuned by the anisotropy and the temperature gradient. Thus, this work not only reproduces the main conclusions obtained in earlier works [Kim \emph{et al}., Phys. Rev. B 92, 020402(R) (2015); Selzer \emph{et al}., Phys. Rev. Lett. 117, 107201 (2016)], but more importantly gives the concrete conditions under which these conclusions apply respectively. Our results may provide useful information on the antiferromagnetic spintronics for future experiments and storage device design.
\end{abstract}


\maketitle

\section{Introduction}
Antiferromangetic (AFM) materials have very recently attracted increasing interest due to their potential applications in the field of antiferromagnetic spintronics\cite{AFMzongshu,2016eleAFM,firstprinciple2017,jia2017}, especially since the modulation and detection of the magnetic state are realized through spin torque effects\cite{90domain,2014XJ} and magnetoresistive properties\cite{2011FMAFM,2014anisotropic,marti2014room,AFMzongshu}. More importantly, the replacement of ferromagnets by antiferromagnets in the active components of spintronic devices is believed to improve the performance stability and to enhance the element density and the information writing/reading speeds of the devices\cite{AFMzongshu,AFMreport}. First, information stored in AFM domains and/or domain walls is insensitive to an applied magnetic field. Second, zero net magnetic moment in an AFM element would not magnetically disturb its neighbors, allowing the elements to be arranged in devices with high density. Furthermore, the frequencies of spin dynamics in an antiferromagnet are much higher than those in a ferromagnet, which makes AFM spintronics more promising. Thus, AFM materials are strongly suggested to play an important role in future storage devices, and several works to search for efficient methods of driving AFM domain wall (DW) motion are available\cite{AFMDWspinwave,AFMDWthermal2016,Mechanical2016,PRL110127208,PRL116147203}.

Most recently, it was experimentally reported that an applied electrical current can trigger the local staggered spin polarizations and in turn induce the local staggered effective field in CuMnAs due to the spin-orbit effect, resulting in the reorientations of antiferromagnetic moments\cite{2016eleAFM,2017moniSOT}. Subsequently, the effective field driven DW motion with a high speed was theoretically demonstrated, suggesting that antiferromagnets such as CuMnAs are good candidates for AFM spintronic applications\cite{gomonay2016high,2016NSOT}. On the other hand, it has been uncovered that various spin wave modes provide different driving forces for the DW motion in an AFM system: linearly polarized spin wave drives DW movtion towards the spin wave source, while circularly polarized one keeps a DW away from the source\cite{AFMDWspinwave,2014AFMSW}. Furthermore, the Dzyaloshinskii-Moriya interaction may results in faster and more controllable motion of AFM-DWs.\cite{arXiv170501572} In the AFM wire system, transverse elastic waves may induce various spin dynamic modes, and the dependence of the spin torque associated with DW on the wave number has been figured out theoretically\cite{Mechanical2016}. However, these predictions are hard to be realized in experiments considering the limitation of technological ability.

Fortunately, thermally driven DW motion under temperature(\emph{T}) gradient in an antiferromagnet has been demonstrated by numerical calculations on a classical spin model in earlier work\cite{AFMDWthermal2016}. On the one hand, similar to the ferromagnetic case\cite{2014ET,2017FMthermal,PRL110177202,PRL111067203}, Selzer \emph{et al} suggested that the entropic torque (ET) induced by the \emph{T}-gradient plays an essential role in the DW motion and drives the wall to the hotter regions. More interestingly, it is suggested that the domain wall is not tilted during its motion, resulting in the absence of the Walker breakdown as well as the lack of inertia in antiferromagnets, which are outstanding merits for technical applications and make antiferromagnets distinctly different from ferromagnets. On the other hand, another recent work by Kim \emph{et al} suggested that the DW in an antiferromagnet should move towards the colder regions driven by a thermal stochastic force caused by the \emph{T}-gradient, based on the fluctuation-dissipation theorem\cite{diwenzou}. This is actually a Brownian motion mode. Therefore, the two consequences as suggested respectively by the two groups are opposite, noting that they studied the same spin model. As a matter of fact, by using an example of a ferromagnetic (FM) DW, the competition between the Brownian force and other \emph{T}-gradient-induced forces such as the entropic force has been discussed in earlier works. It is suggested that the DW width is very important in modulating these forces and affects the motion of the FM-DW\cite{2015FMKim}. These important works afford us valuable experience in the study of AFM-DW motion under a \emph{T}-gradient. This study is very important for basic physics due to the fact that the \emph{T}-gradient driven domain wall motion may apply to other modulation methods such as N$\acute{e}$el spin-orbit torques\cite{2014first,2016eleAFM,2017SOT}. Moreover, it could provide detailed information on thermally driven DW motion, allowing one to control this motion accurately through elaborately adjusting process parameters, which is very meaningful for future spintronic device design. Thus, this study is essential both in basic physical research and for potential applications.

In this work, we study the AFM spin dynamics of a classical spin model with an uniaxial anisotropy under a \emph{T}-gradient, using the stochastic Landau-Lifshitz-Gilbert simulations. We figure out that the entropic torque driven motion and the Brownian motion of the DWs always coexist in an anitiferromagnet and a strong competition between them is expected. The comprehensive effect is mainly determined by the magnitudes of the uniaxial anisotropy and the \emph{T}-gradient. Therefore, the present work not only reproduces well the main conclusions reached in the earlier works\cite{AFMDWthermal2016,diwenzou}, but also gives the concrete conditions under which these conclusions apply respectively. Our results appear to be a more general approach to the AFM spin dynamics.

The remaining part of this paper is organized as follows: In Sec. II, the model and the computation method will be described. Section III is devoted to the numerical results and discussion, and the conclusion is presented in Sec. IV.

\section{Model and method}

\begin{figure}[tbp]
\centering\includegraphics[width=8.6cm]{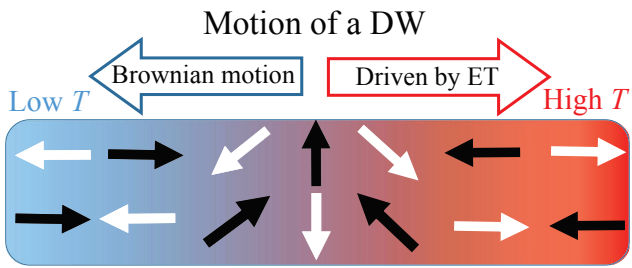}
\caption{The entropic torque driven domain wall motion and the Brownian motion of an AFM domain wall under a temperature gradient.}
\label{fig1}
\end{figure}

We study the dynamics of an AFM-DW under a linear \emph{T}-gradient based on a classical spin model with isotropic Heisenberg exchanges between the nearest neighbors and an uniaxial anisotropy term. The model Hamiltonian is stated as
\begin{equation}
\label{eq.1}
\emph{H}=-\emph{J}\sum_{<\emph{i},\emph{j}>}\textbf{S}_{i}\cdot\textbf{S}_{j}-
\emph{D}_{\emph{z}}\sum_{\emph{i}}(\emph{S}^{z}_{i})^{2}
\end{equation}
where \emph{J} $<$ 0 is the AFM coupling magnitude, \textbf{S}$_{\emph{i}}$ represents the normalized magnetic moment at site \emph{i}, and \emph{D}$_{\emph{z}}$ $>$ 0 is the anisotropy constant defining an easy axis in the \emph{z}  (wire axis) direction. The AFM spin dynamics at finite \emph{T} is investigated by the stochastic Landau-Lifshitz-Gilbert equation\cite{LLG1,LLG2,atomLLG},
\begin{equation}
\label{eq.2}
\frac{\partial \textbf{S}_{\mathit{i}}}{\partial \mathit{t}} = \frac{\gamma }{\mu _{\textit{s}}(1 + \alpha ^{2})}\textbf{S}_{\mathit{i}} \times [\textbf{H}_{\mathit{i}} - \alpha (\textbf{S}_{\mathit{i}} \times \textbf{H}_{\mathit{i}})]
\end{equation}
where $\gamma$ is the gyromagnetic ratio, $\alpha\equiv$ 0.01 is the Gilbert damping constant\cite{AFMDWthermal2016,damping}, \textbf{H}$_{i}$ = -$\partial$\emph{H}/$\partial$\textbf{S}$_{i}$ + $\zeta_{i}$(\emph{t}) is the effective field with an additional white noise term representing the thermal fluctuations (random field) satisfying the
fuctuation-dissipation theorem: $\langle$$\zeta_{i}$(\emph{t})$\rangle$ = 0 and $\langle$$\zeta_{i\eta}$(\emph{t})$\zeta_{j\theta}$(\emph{t}$'$)$\rangle$ = $\delta_{i,j}$$\delta_{\eta,\theta}\delta$(\emph{t} - \emph{t}$'$)2$\alpha$\emph{k}$_{B}$\emph{T}$\mu_{s}$/$\gamma$\cite{1963,classicalSpin}. The Landau-Lifshitz-Gilbert simulations are performed on an AFM nanowire described as a \emph{w} $\times$ \emph{h} $\times$ \emph{L} (\emph{w} = 6, \emph{h} = 6, \emph{L} = 600 unless stated elsewhere) elongated three-dimensional lattice (lattice parameter a$_{0}$) with open boundary conditions using the Heun¡¯s method with a time step $\Delta$\emph{t}=2 $\times$ 10$^{-4}|\mu_{s}$/$\gamma$$J|$\cite{classicalSpin}. After sufficient relaxation for a DW, we apply a linear \emph{T}-gradient to the lattice along the \emph{z}-axis and observe the AFM-DW motion. As long as the temperature gradient is sufficiently small, equilibrium thermodynamics can be applied.Especially, various \emph{D}$_{\emph{z}}$ and \emph{T}-gradients $\nabla$\emph{T} are studied in details, and the remaining parameters are fixed for simplicity.

Furthermore, Monte Carlo (MC) simulations are also performed using the standard Metropolis algorithm\cite{MC} to obtain the equilibrium state of the lattice and to calculate some thermodynamic parameters. Specifically, the staggered magnetization is defined as \emph{n}(\emph{T}) = $|$\textbf{m}$_{1}$(\emph{T}) - \textbf{m}$_{2}$(\emph{T})$|$, where \textbf{m}$_{1,2}$ are the magnetizations of the two sublattices. Furthermore, the internal energy \emph{E} = $\langle$\emph{H}$\rangle$ and its fluctuation $\delta$\emph{E} = $\langle$\emph{H}$^{2}\rangle$ - $\langle$\emph{H}$\rangle^{2}$ are also calculated in order to calculate the free energy fluctuation $\delta$\emph{f} through the free energy $\Gamma$ vs. \emph{E} relation
\begin{equation}
\label{eq.4}
\Gamma(\beta)=\frac{1}{\beta}\int_{0}^{\beta}\mathit{E}(\beta')\mathit{d}\beta '
\end{equation}
with $\beta$ = 1/\emph{k}$_{\emph{B}}$\emph{T}.

\section{Numerical results and discussion}
\subsection{Entropic torque driven domain wall motion}
\begin{figure}[tbp]
\centering\includegraphics[width=8cm]{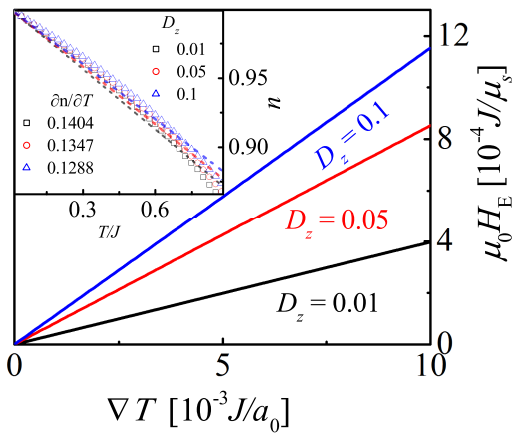}
\caption{Entropic field as a function of temperature gradient $\nabla$\emph{T} for various \emph{D}$_{z}$. The inset shows the N$\acute{e}$el vector as a function of temperature for various \emph{D}$_{z}$}
\label{fig2}
\end{figure}
First, we study the entropic torque driven DW motion when a \emph{T}-gradient is applied. Here, this effective torque is caused by the thermodynamic effect of minimizing the free energy, completely different from the well-known spin transfer torque which is normally resulted from direct interaction between spin current and local spin. According to the theory on entropic torque, a \emph{T}-gradient in an antiferromagnet induces a DW energy gradient and in turn generates an equivalent entropic field\cite{2014ET,AFMDWthermal2016,2017FMthermal}
\begin{equation}
\label{eq.3}
\mu_{0}\mathbf{H} _{E} = -\frac{2\emph{a}_{0}^{2}\emph{J}}{\mu_{s}\Delta_{0}} \frac{\partial\emph{n}}{\partial\emph{T}}\nabla\emph{T}
\end{equation}
where $\Delta_{0}$ = $\pi$\emph{a}$_{0}\sqrt{\emph{J}/2\emph{D}_{z}}$ is the DW width($\sim$ 7a$_{0}$ for \emph{D}$_{z}$ = 0.1\emph{J}), the average N$\acute{e}$el vector \emph{n} is temperature dependent and obtained from the MC simulations. It is clearly shown in Eq. (4) that the entropic field is significantly dependent on the magnitude of the uniaxial anisotropy. The MC simulated \emph{n}(\emph{T}) curves for various \emph{D}$_{\emph{z}}$ are given in the insert of Fig. 2. In the low \emph{T} region, the \emph{n}(\emph{T}) curve can be linearly fitted, and $\partial$\emph{n}$/\partial$\emph{T} is reasonably calculated. With the increase of \emph{D}$_{\emph{z}}$, the anisotropy energy is enhanced, further stabilizing the AFM order ($\partial$\emph{n}/$\partial$\emph{T} is decreased). Furthermore, $\Delta_{0}$ is inversely proportional to $\sqrt{\emph{D}_{z}}$, and significantly deceases as \emph{D}$_{\emph{z}}$ increases. As a result, for a fixed \emph{T}-gradient, the entropic field increases with the increasing uniaxial anisotropy, as clearly shown in Fig. 2 which presents the calculated field as a function of $\nabla$\emph{T} for various \emph{D}$_{\emph{z}}$. In this ideal case, the field simultaneously increases from zero when $\nabla$\emph{T} is applied, resulting in the DW motion to the hotter regions, as revealed in the work of Selzer \emph{et al}\cite{AFMDWthermal2016}. This physical process can be well explained from the free energy landscape, as schematically depicted in Fig. 3(a). During the DW motion, the entropy of the system is increased (\emph{S} $\rightarrow$ \emph{S}$'$), while the free energy is decreased (\emph{F} $\rightarrow$ \emph{F}$'$). Thus, even a small entropic torque (\emph{F} $\approx$ \emph{F}$'$) is sufficient to drive the DW motion.

\begin{figure}[tbp]
\centering\includegraphics[width=8.5cm]{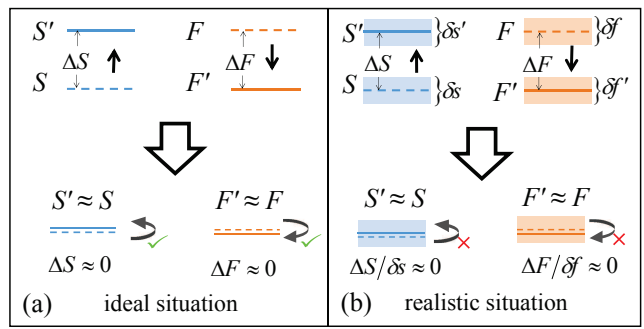}
\caption{Schematic of entropy(free energy) transition during the entropic torque driven domain wall motion. (a)in the ideal case and (b) in the realistic case considering thermal fluctuations}
\label{fig2}
\end{figure}

However, thermal fluctuations cannot be avoided at finite \emph{T}, and the fluctuations of entropy ($\delta$\emph{s} and $\delta$\emph{s}$'$) and free energy ($\delta$\emph{f} and $\delta$\emph{f}$'$) are inevitable, as depicted in Fig. 3(b). Thus, when the free energy difference $\Delta$\emph{F} = \emph{F} - \emph{F}$'$ is much smaller than the free energy fluctuation, the entropic torque driven DW motion could be suppressed. In order to study the free energy difference, we calculate the free energy of DW as a function of \emph{T} by\cite{2017FMthermal,FM2008MFA}
\begin{equation}
\emph{e} = \frac{2\sqrt{2\emph{D}_{z}\emph{J}}}{\emph{a}_{0}^{2}}\emph{n}(\emph{T})^{2}.
\end{equation}
Fig. 4(a) shows the calculated \emph{e}(\emph{T}) curves for various \emph{D}$_{\emph{z}}$. With decreasing \emph{D}$_{\emph{z}}$, \emph{e} is obviously decreased. Furthermore, we also calculate the ratio of $\Delta$\emph{F} to $\delta$\emph{f}, i.e. $\rho$ = $\Delta$\emph{F}/$\delta$\emph{f}, to qualitatively describe the effect of entropic torque. Without loss of generality, $\Delta$\emph{F} is calculated by $\Delta$\emph{F}(\emph{T}$_{1}$,\emph{T}$_{2}$) = \emph{S}$_{c}$$|$\emph{e}(\emph{T}$_{2}$) -  \emph{e}(\emph{T}$_{1}$)$|$  where \emph{S}$_{c}$ is the cross-sectional area of the nanowire, and $\delta$\emph{f} is the free energy fluctuation at the average temperature $\delta$\emph{f}((\emph{T}$_{1}$ + \emph{T}$_{2}$)/2) with \emph{T}$_{1}$ and \emph{T}$_{2}$ are temperatures at the two ends of the nanowire lattice. Fig. 4(b) gives the calculated $\rho$ in the (\emph{T}$_{1}$, \emph{T}$_{2}$) parameter plane at \emph{D}$_{\emph{z}}$ = 0.1\emph{J} for \emph{L} = 200. It is clearly shown that $\rho$ quickly increases as the \emph{T}-gradient increases, consistent with Eq. (3). Thus, given the same temperature difference, $\rho$ is significantly decreased for a larger \emph{L}, as shown in Fig. 4(c) which gives the calculated $\rho$ for \emph{L} = 400. Furthermore, the calculated $\rho$ at \emph{D}$_{\emph{z}}$ = 0.01 for \emph{L} = 400 is presented in Fig. 4(d) which demonstrates an important dependence of $\rho$ on \emph{D}$_{\emph{z}}$. It is noted that \emph{e} is remarkably reduced with decreasing \emph{D}$_{\emph{z}}$, resulting in the decrease of $\rho$. As a result, for an AFM system with small \emph{D}$_{\emph{z}}$ under weak \emph{T}-gradient, the effect of entropic torque should be significantly suppressed, resulting in the suppression of the entropic torque driven DW motion to the hotter regions.

\begin{figure}[tbp]
\centering\includegraphics[width=8.5cm]{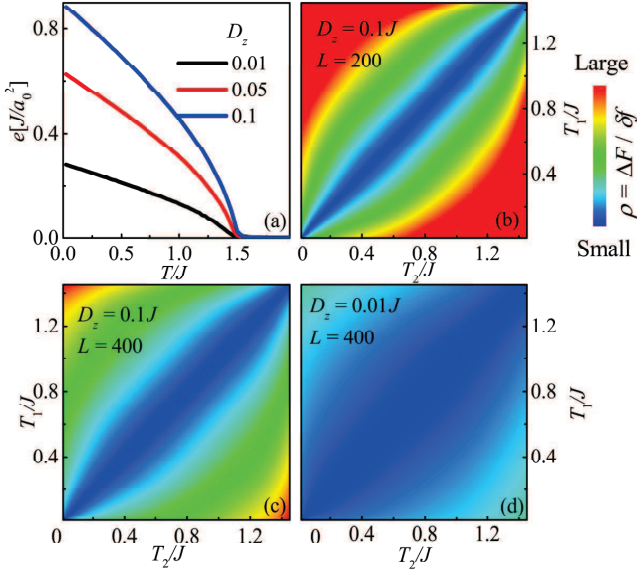}
\caption{(a)The calculated domain wall free energy as a function of temperature for various \emph{D}$_{z}$.The calculated $\rho$ in the (\emph{T}$_{1}$, \emph{T}$_{2}$) for (b)\emph{D}$_{z}$ = 0.1\emph{J}, \emph{L} = 200, and  (c)\emph{D}$_{z}$ = 0.1\emph{J}, \emph{L} = 400, and (d)\emph{D}$_{z}$ = 0.01\emph{J}, \emph{L} = 400. \emph{T}$_{1}$ and \emph{T}$_{2}$ are temperatures at the two ends of lattice along the \emph{z}-axis.}
\label{fig3}
\end{figure}

\subsection{Brownian motion of a domain wall}
In this subsection, we study the Brownian motion of DW in the absence of \emph{T}-gradient using the Landau-Lifshitz-Gilbert simulations performed on an 8 $\times$ 8 $\times$ 80 lattice with periodic boundary condition which is applied in the wire axis direction in order to maintain the DW. Fig. 5 presents the deviation of the DW position from the midpoint of the lattice $\Delta$\emph{L} as a function of time $\tau$ for various \emph{T}. It is clearly shown that the DW is strongly deviated from the midpoint by thermal fluctuations at high \emph{T} (\emph{T} $\geq$ 0.01\emph{J}). With the decreasing \emph{T}, thermal fluctuations and the midpoint deviation of the DW are quickly suppressed. At low \emph{T} (\emph{T} $<$ 10$^{-4}$\emph{J}), the deviation is completely diminished and the DW hardly moves, as clearly shown in Fig. 5, an obvious fact.

To some extent, the DW shifting due to the thermal fluctuations at finite \emph{T} can be analogically understood based on the classical particle Brownian motion theory. Thus, we further study the probability distribution of the DW positions (PDP) and give the results in Fig. 6. Here, over 1200 samples are simulated and counted in order to achieve reasonable data reliability. It is clearly shown that all PDP curves are axis-symmetric ($\Delta$\emph{L} = 0) and exhibit the character of the Gaussian distribution, well consistent with the conclusion of the classical Brownian motion theory\cite{1930BM,1985BM}. Furthermore, the width of the Gaussian function is increased with increasing time, as clearly shown in Fig. 6(a) which presents the results at \emph{T} = 0.1\emph{J} and \emph{D}$_{\emph{z}}$ = 0.1\emph{J} for various times, further confirming the Brownian motion nature of DW. On the other hand, Fig. 6(b) shows the calculated PDP at \emph{T} = 0.1\emph{J} and $\Delta\tau$ = 50 for various \emph{D}$_{\emph{z}}$. For a fixed shifting time, the width of the Gaussian function increases with the decrease of \emph{D}$_{\emph{z}}$, demonstrating that the Brownian motion of an AFM-DW is not favored by the uniaxial anisotropy. Thus, in the Brownian motion of DW, \emph{D}$_{\emph{z}}$ plays a similar role as that of the particle mass (inertia) in the classical theory. As a result, the Brownian motion of DW is enhanced for a system with small \emph{D}$_{\emph{z}}$, contrary to the DW motion driven by the entropic torque.

Furthermore, it is well known that the particle probability density in the colder region is larger than that in the hot region due to the different mobilities of the particles under an applied \emph{T}-gradient. Similarly, the Brownian motion of an AFM-DW is also a statistical behavior, ensuring that one observes more likely the DW in the low \emph{T} regions. Thus, the Brownian motion effect and entropic torque effect for an AFM-DW compete with each other, and the motions can be modulated by the uniaxial anisotropy and the applied \emph{T}-gradient, to be discussed in detail in the following subsection.

\begin{figure}[tbp]
\centering\includegraphics[width=8cm]{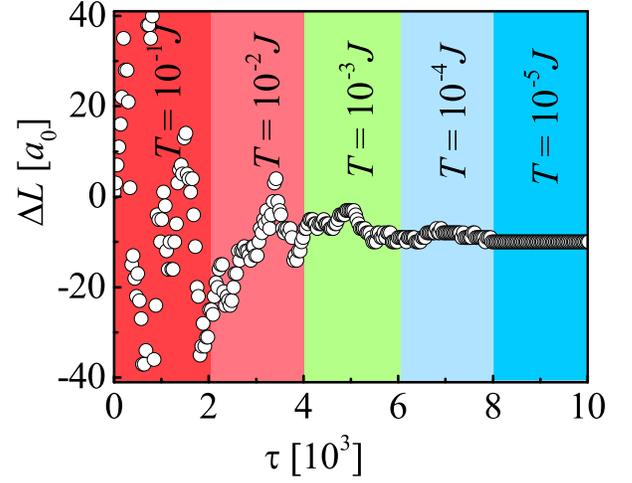}
\caption{Displacement $\Delta$\emph{L} for a domain wall versus time at \emph{D}$_{z}$ = 0.1\emph{J} and system cools down after every 2000 $\tau$ from 10$^{-1}$\emph{J} to 10$^{-5}$\emph{J}}
\label{fig4}
\end{figure}

\begin{figure}[tbp]
\centering\includegraphics[width=8.5cm]{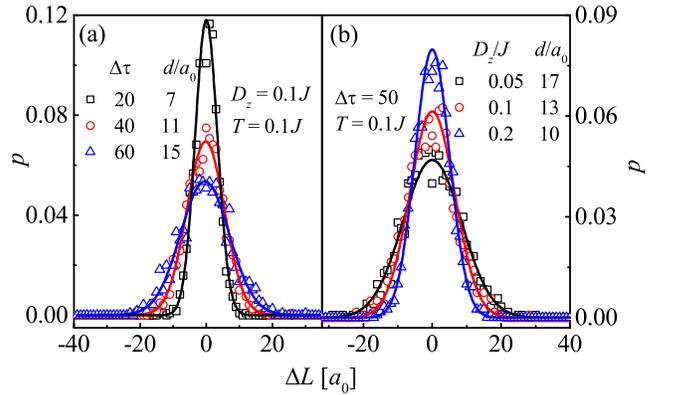}
\caption{The probability distribution of position \emph{p} at \emph{T} = 0.1\emph{J} without \emph{T}-gradient (a) for various elapsed time at \emph{D}$_{z}$ = 0.1\emph{J}, and (b) for various \emph{D}$_{z}$ at $\tau$ = 200. }
\label{fig6}
\end{figure}

\subsection{Competition between Brownian motion and entropic torque}
First, we highlight in Table 1 the effects of the uniaxial anisotropy and \emph{T}-gradient on the DW motion driven respectively or together by the Browaian motion and entropic torque, as characterized by the driving force and speed, as uncovered in the earlier works\cite{AFMDWthermal2016,diwenzou} and in this work. As a matter of fact, these effects have been confirmed in our Landau-Lifshitz-Gilbert simulations. On the one hand, Fig. 7(a) shows the deviation of the DW from the  initial point for various \emph{D}$_{z}$ at $\nabla$\emph{T} = 8.3 $\times$ 10$^{-6}$ \emph{J}/\emph{a}$_{0}$. For small \emph{D}$_{z}$ = 0.01, the DW moves toward the colder region which is mainly attributed to the Brownin motion. It is noted from Table 1 that \emph{D}$_{z}$ hinders the Brownian motion of the DW and enhances the entropic torque, which speeds down the wall motion. Thus, the DW motion becomes slower with increasing \emph{D}$_{z}$ and even reverses for large enough \emph{D}$_{z}$ (\emph{D}$_{z}$ $>$ 0.1), as clearly shown in our simulated results.

\renewcommand\arraystretch{1.5}
\begin{table}[]
\centering
\caption{Influence of different parameters on the driving force(\emph{F}$_{d}$) or speed(\emph{V}) of the entropic torque (ET) and Brownian motion (BM) refer to [\cite{AFMDWthermal2016}] and [\cite{diwenzou}] where "+" is for positive effect, "$-$" for negetive and "=" for ineffectiveness.}
\label{my-label}
\begin{tabular}{p{1.5cm}p{1.5cm}p{1.5cm}p{1.5cm}}
\hline
                                         &            & \emph{D}$_{z}$ & $\nabla$\emph{T} \\ \hline
\multicolumn{1}{c|}{\multirow{2}{*}{ET}} & \textit{F}$_{d}$ & +       & +                \\
\multicolumn{1}{c|}{}                    & \textit{V} & $\times$                & +                \\ \hline
\multicolumn{1}{c|}{\multirow{2}{*}{BM}} & \textit{F}$_{d}$ & $\times$                & +                         \\
\multicolumn{1}{c|}{}                    & \textit{V} & $-$     & +                          \\ \hline
\end{tabular}
\end{table}

\begin{figure}[tbp]
\centering\includegraphics[width=8.5cm]{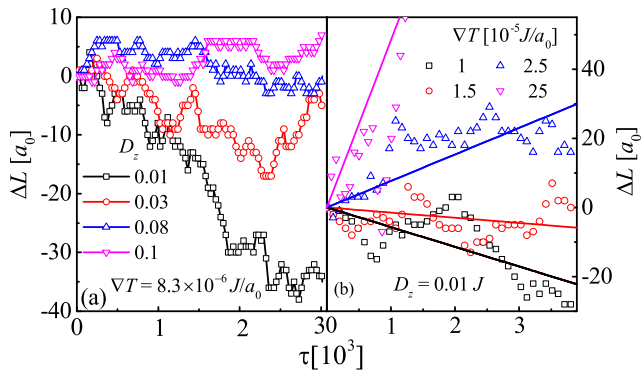}
\caption{Displacement of the domain wall versus time (a) for various \emph{D}$_{z}$ at $\nabla$\emph{T} = 8.3 $\times$ 10$^{-6}$\emph{J}/\emph{a}$_{0}$ and (b) for various $\nabla$\emph{T} at \emph{D}$_{z}$ = 0.01\emph{J}.}
\label{fig7}
\end{figure}
On the other hand, an applied \emph{T}-gradient contribute both to the Brownian motion and entropic torque driven motion of the DW, resulting in a strong competition between the two motions. Thus, the Landau-Lifshitz-Gilbert simulations are also performed to make clearly the comprehensive effects of the \emph{T}-gradient, and the corresponding results are shown in Fig. 7(b). For small $\nabla$\emph{T} = 1 $\times$ 10$^{-5}$ \emph{J}/\emph{a}$_{0}$ at \emph{D}$_{z}$ = 0.01, the DW apparently moves toward the colder region. Similarly, the velocity significantly decreases with the increase of $\nabla$\emph{T}, suggesting that the \emph{T}-gradient favors the entropic torque driven motion more than the Brownian motion of the DW. When $\nabla$\emph{T} increases to above 2 $\times$ 10$^{-5}$ \emph{J}/\emph{a}$_{0}$, the entropic torque driven motion takes advantage of the Brownian motion, resulting in the motion of the DW toward the hotter regions, as uncovered in our simulations.

In earlier works, it has been pointed out that for the FM/AFM case that a Brownian force dominates the other forces when the DW width is large enough, and the other forces such as an entropic force dominate a Brownian force when the DW width is small\cite{diwenzou,2015FMKim}. As a matter of fact, the numerical results of the AFM-DW motion in this work also agree well with the analytical result. Specifically, larger/smaller \emph{D}$_{\emph{z}}$ yields a smaller/larger DW width, which makes an entropic/Brownian force stronger. Thus, the present work seems to reveal once more that spin dynamics of antiferromagents are complex and fascinating, and can be well modulated through tuning related parameters. The motions of the AFM-DW under applied \emph{T}-gradients have attracted attention for a few years, but their microscopic mechanisms had been ambiguous before the present simulations.

\section{Summary}
In summary, we have studied the spin dynamics in antiferromagnetic nanowire under the applied temperature gradient using micromagnetic simulations in combination of Monte Carlo simulations. It is revealed that there is a strong competition between the Brownian motion and the entropic torque driven motion of the domain-wall, resulting in a complex spin dynamics in antiferromagnets. Furthermore, the uniaxial anisotropy and the temperature gradient have been confirmed to play essential roles in determining the motion of the wall. Thus, the motions of antiferromagnetic domain-wall in opposite directions have been well explained in this work, which may provide useful information for future storage device design.

\section{Acknowledgement}
We are grateful for insightful discussions with Se Kwon Kim, Xingtao Jia, Chenguang Yang, Ping Tang, Liu Huo and Xiufeng Han. The work is supported by the National Key Projects for Basic Research of China (Grant No. 2015CB921202), and the National Key Research Programme of China (Grant No. 2016YFA0300101), and the Natural Science Foundation of China (Grant No. 51332007), and the Science and Technology Planning Project of Guangdong Province (Grant No. 2015B090927006), and Special Funds for Cultivation of Guangdong College Students¡¯ Scientific and Technological Innovation (Grant No. pdjh2017b0138). X. Lu also thanks for the support from the project for Guangdong Province Universities and Colleges Pearl River Scholar Funded Scheme (2016).
\bibliographystyle{apsrev}
\bibliography{AFMdomain}

\end{document}